\documentclass{webofc}

\usepackage[utf8]{inputenc} 
\usepackage[T1]{fontenc}    
\usepackage{rotating}
\usepackage{array}
\usepackage{amsmath}
\usepackage[normalem]{ulem}
\usepackage{slashed}
\usepackage{booktabs}
\usepackage[pdftex,table]{xcolor}
\usepackage{units}
\usepackage{xfrac}
\usepackage{mathtools}
\usepackage{empheq}
\usepackage[]{units}
\usepackage{multirow}
\usepackage{amssymb}
\usepackage{url}
\usepackage{comment}
\usepackage{paralist}
\usepackage{xspace}
\usepackage{hyperref}

\usepackage{tikz}

\def\beq{\begin{equation}}
\def\eeq{\end{equation}}
\newcommand{\bea}{\begin{eqnarray}\begin{aligned}}
\newcommand{\eea}{\end{aligned}\end{eqnarray}}
\def\bitem{\begin{itemize}}
\def\eitem{\end{itemize}}

\newcommand{\pt}{\ensuremath{p_\text{T}}\xspace}
\newcommand{\pip}{\ensuremath{\pi^-p}\xspace}

\begin{document}

\title{Simulation of Hadronic Interactions with Deep Generative Models}

\author{
\firstname{Tuan Minh} \lastname{Pham}\inst{1}\fnsep\thanks{\email{tuan.minh.pham@cern.ch}} 
\and
\firstname{Xiangyang} \lastname{Ju}\inst{2}\fnsep\thanks{\email{xju@lbl.gov}} 
}

\institute{
Physics Department, University of Wisconsin-Madison, Madison, WI 53706
\and
Scientific Data Division, Lawrence Berkeley National Laboratory, Berkeley, CA 94720
}

\abstract{
Accurate simulation of detector responses to hadrons is paramount for all physics programs at the Large Hadron Collider (LHC). Central to this simulation is the modeling of hadronic interactions. Unfortunately, the absence of first-principle theoretical guidance has made this a formidable challenge. The state-of-the-art simulation tool, \textsc{Geant4}, currently relies on phenomenology-inspired parametric models. Each model is designed to simulate hadronic interactions within specific energy ranges and for particular types of hadrons. Despite dedicated tuning efforts, these models sometimes fail to describe the data in certain physics processes accurately.
Furthermore, fine-tuning these models with new measurements is laborious. Our research endeavors to leverage generative models to simulate hadronic interactions. While our ultimate goal is to train a generative model using experimental data, we have taken a crucial step by training conditional normalizing flow models with \textsc{Geant4} simulation data. Our work marks a significant stride toward developing a fully differentiable and data-driven model for hadronic interactions in High Energy and Nuclear Physics.
}

\maketitle

\section{Introduction} \label{sec:intro}

Hadronic interactions exhibit a well-defined description in high-energy domains, typically in the hundreds of GeV range, where perturbative Quantum Chromodynamics (QCD) proves effective. However, in lower energy regimes, the perturbative theory loses its applicability. To address this challenge, various phenomenological models have been devised to characterize hadronic interactions. These models are designed for specific kinematic regions and are tailored to a limited number of hadron flavors. To provide a comprehensive description across a wide energy spectrum, spanning from MeV to hundreds of GeV, these models must be integrated into a transport code. This integration is precisely achieved in the \textsc{Geant4} framework~\cite{GEANT4:2002zbu}. It serves as a unified platform where these diverse models are combined to simulate hadronic interactions across an extensive energy range.

The \textsc{Geant4} framework plays a vital role in simulating detector effects for a broad spectrum of applications, including the Large Hadron Collider experiments and beyond. The \textsc{Geant4} simulation, often referred to as ``full simulation'', constitutes the most computationally intensive aspect of collider physics programs. As detectors are progressively enhanced to achieve finer resolution, the computational demands of full simulation are poised to escalate significantly.

To address this challenge, a concerted effort is emerged towards the development of fast simulation techniques employing deep generative models~\cite{1701.05927,1705.02355,1712.10321,ATLAS:2021pzo,ATL-SOFT-PUB-2020-006,ATL-SOFT-PUB-2018-001,Chekalina:2018hxi,Deja:2019vcv,barin_pacela_vitoria_2018_1470512}, despite that much improvement has been made within the \textsc{Geant4} framework~\cite{Dotti:2011zz}. Recent studies have shown that Normalizing Flow~\cite{pmlr-v37-rezende15} can achieve state-of-the-art precision while dramatically reducing generation time, yielding computational gains of orders of magnitude compared to full simulation~\cite{Krause:2021wez,Krause:2021ilc}. However, these generative models are tailored to high-level detector-specific features, making them unsuitable for generalized use across different particle detectors. Our ultimate objective is to develop a generative model that learns the non-perturbative hadronic interactions from the experimental data, such as those published in Refs.~\cite{E-802:1991unu,MIPP:2010vnr}. 


\section{Methods} \label{sec:methods}

\subsection{Dataset} \label{subsec:dataset}

The data is generated by the \textsc{GEANT4}~\cite{GEANT4:2002zbu} toolkit with a physics list that includes \textsc{Bertini} Cascade and \textsc{Fritiof} model (a.k.a \textsc{FTFP\_BERT\_ATL}). The \textsc{Bertini} model is applicable for incident energies below 10 GeV in most use cases, while the \textsc{Fritiof} model is for incident energies from 9 GeV to 100 TeV. For simplicity, we use a uniform energy transition for the \textsc{FTFP\_BERT\_ATL} physics list. 

\begin{figure}[tbh]
    \centering
    \includegraphics[width=0.5\textwidth]{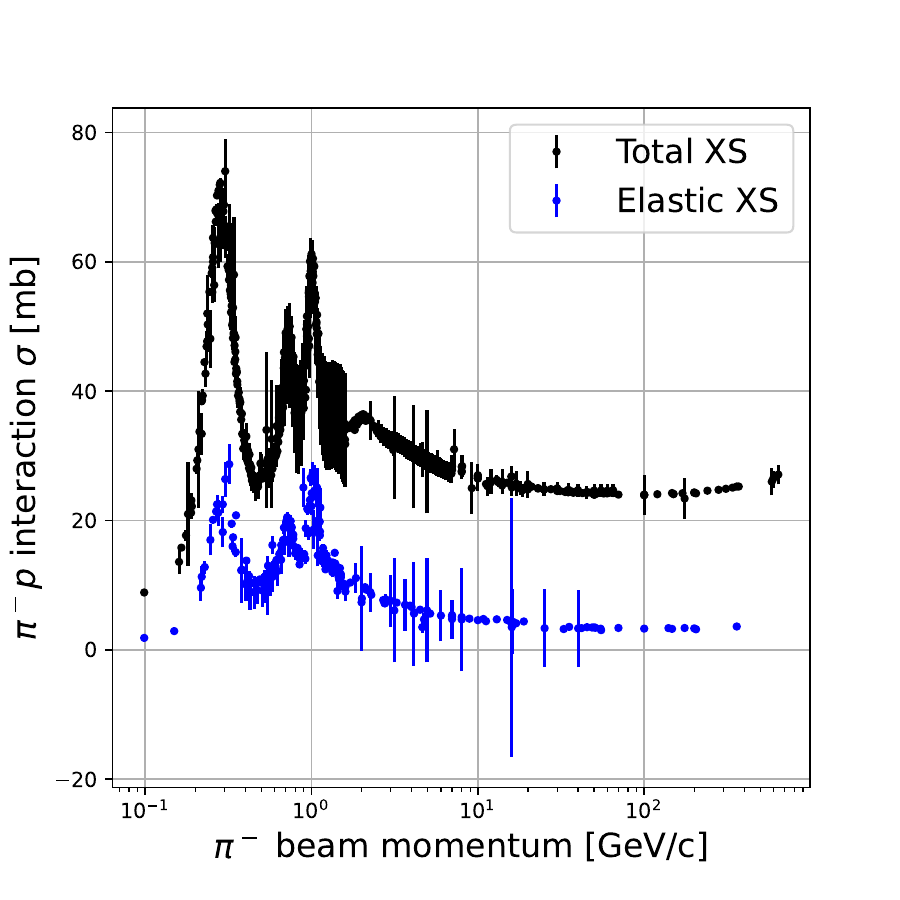}
    \caption{Total and elastic cross section of \pip interactions as a function of incident $\pi^-$ kinematic energy in the lab frame, taken from PDG data-base~\cite{pdg}.}
    \label{fig:pion-proton-tot-xs}
\end{figure}

We chose the \pip interactions for our studies, whose total cross sections are measured across a large range of kinetic ranges as shown in Fig.~\ref{fig:pion-proton-tot-xs}. There are peaks in the cross sections connected with the $\Delta$-isobar production in the $s$-channel, $\pi^- + p \to \Delta^0$. The main decay mode of a $\Delta^0$ is $\Delta^0 \to \pi^- + p$. Thus, we focus on simulating events with two outgoing hadrons. 

To span a broad spectrum of incident $\pi^-$ energies, we selected 29 data points between 100 MeV and 8000 MeV, spaced approximately 200 MeV apart for training. Additionally, we employed 14 distinct energy levels within the range of 200 MeV to 6500 MeV for testing. For each energy point, we generated 100,000 events with randomly sampled momentum directions. In summary, we produced 290,000 training events and 140,000 testing events.

\begin{figure}[htb]
    \centering
    \includegraphics[width=0.49\textwidth]{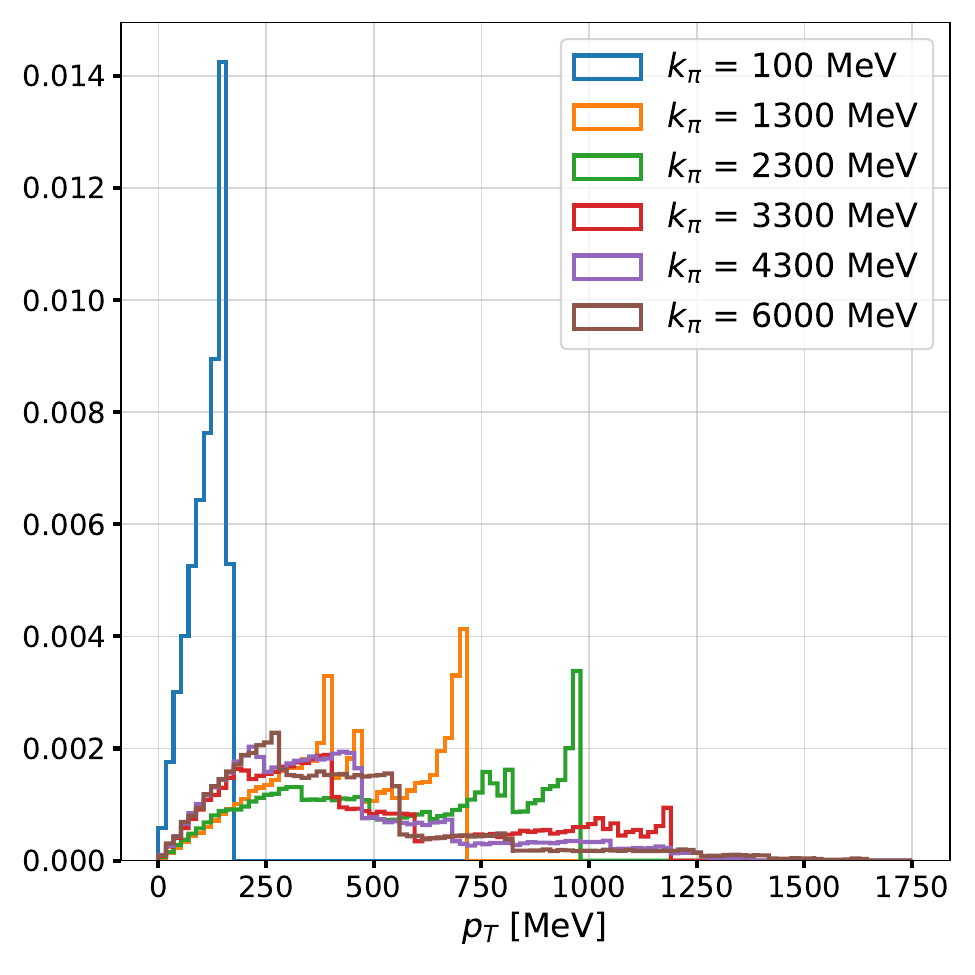}
    \includegraphics[width=0.49\textwidth]{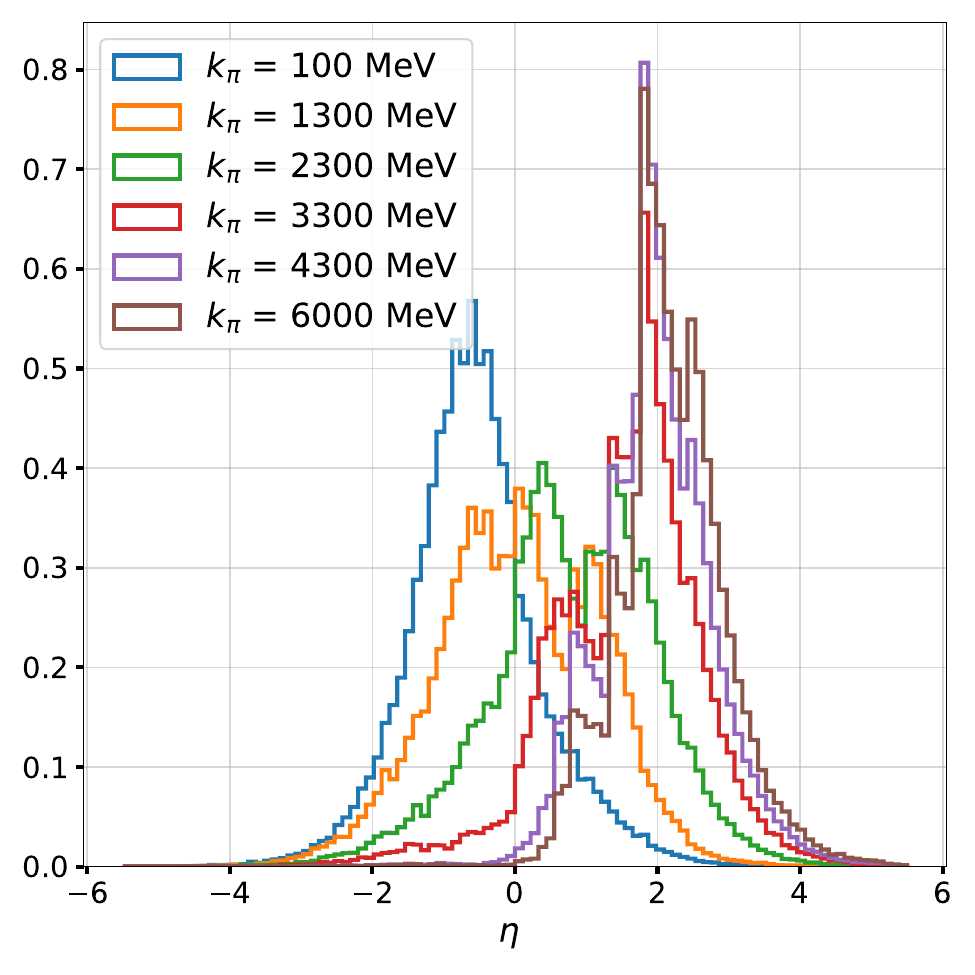}
    \caption{Comparison of the transverse momentum \pt and pseudorapidity $\eta$ distributions of the leading final state particle for different kinematic energy of the incoming $\pi^-$. Final state particles are sorted by their \pt. }
    \label{fig:low-energy-distribution}
\end{figure}

At the center-of-mass (COM) frame, the two outgoing particles are produced back-to-back, distributed uniformly across the azimuthal angle. Thanks to energy conservation, our generative model only needs to predict the properties of one of these particles. Figure~\ref{fig:low-energy-distribution} shows the transverse momentum and pseudorapidity of the leading particle in events featuring two outgoing particles. The kinetic energies of the leading particles change dramatically when the incident kinetic energy changes, attributed to multiple physics processes. These dependencies gradually smooth out when the pion energy exceeds 10~GeV, at which point only the \textsc{Fritiof} model is employed. Accurately modeling the correlations between the leading particle energy and the incident kinetic energy poses a formidable challenge. 

\subsection{Normalizing Flow and Training} \label{subsec:nftraing}
A normalizing flow (NF) uses an invertible function $f$ (also known as a bijector) to transform a simple initial density $\pi(\vec{z})$ to the target density distribution $p(\vec{x})$, and the autoregressive density estimator models any joint density $p(\vec{x})$ as a product of one-dimensional conditional distributions. We use a simple invertible function $f$ in the MAF: $x_i = f(z_i) = z_i \exp(\alpha_i) + \beta_i$ where $\alpha_i$ and $\beta_i$ are parameterized by neural networks, often by the MultiLayer Perceptrons (MLPs). The Adam optimizer then optimizes the learnable weights in the neural network by minimizing the negative log-likelihood function. A normalizing flow can be extended to a conditional normalizing flow by concatenating the conditional vector $\vec{c}$ with the input vector $\vec{x}$ and using the combined vector to estimate the target density distribution.

Our study employs a specific variant of normalizing flows called Masked Autoregressive Flow (MAF)~\cite{PapamakariosMAF}. MAF constructs multi-dimensional distributions by sequentially modeling each dimension based on previously modeled ones. While this autoregressive approach enhances modeling accuracy, it can introduce sampling bottlenecks and sensitivity to the input vector's order. To mitigate the ordering impact, we introduced a permutation bijection to each MAF block. Additionally, in the final block, we appended a $\tanh$ bijector layer to ensure that the outputs fall within the target distribution's range, namely [-1, 1].

Our final normalizing flow model uses the incident pion energy as the conditional variable and employs Gaussian distributions as the base distributions. Its target distributions are the four-vector of leading outgoing particles ($p_x$, $p_y$, $p_z$, and $E$). The normalizing flow consists of 30 Masked Autoregressive Flow (MAF) blocks, with each MAF block using two-layer MLPs with a layer size of 128.  The implementation is based on \textsc{TensorFlow}~\cite{tensorflow2015-whitepaper} (TF) and \textsc{TF probability}. The model undergoes training for over 2000 epochs with a learning rate that decreases from $10^{-4}$ to $10^{-6}$ following a polynomial function. 

\section{Results}

\begin{figure}
    \centering
    \includegraphics[width=0.6\textwidth]{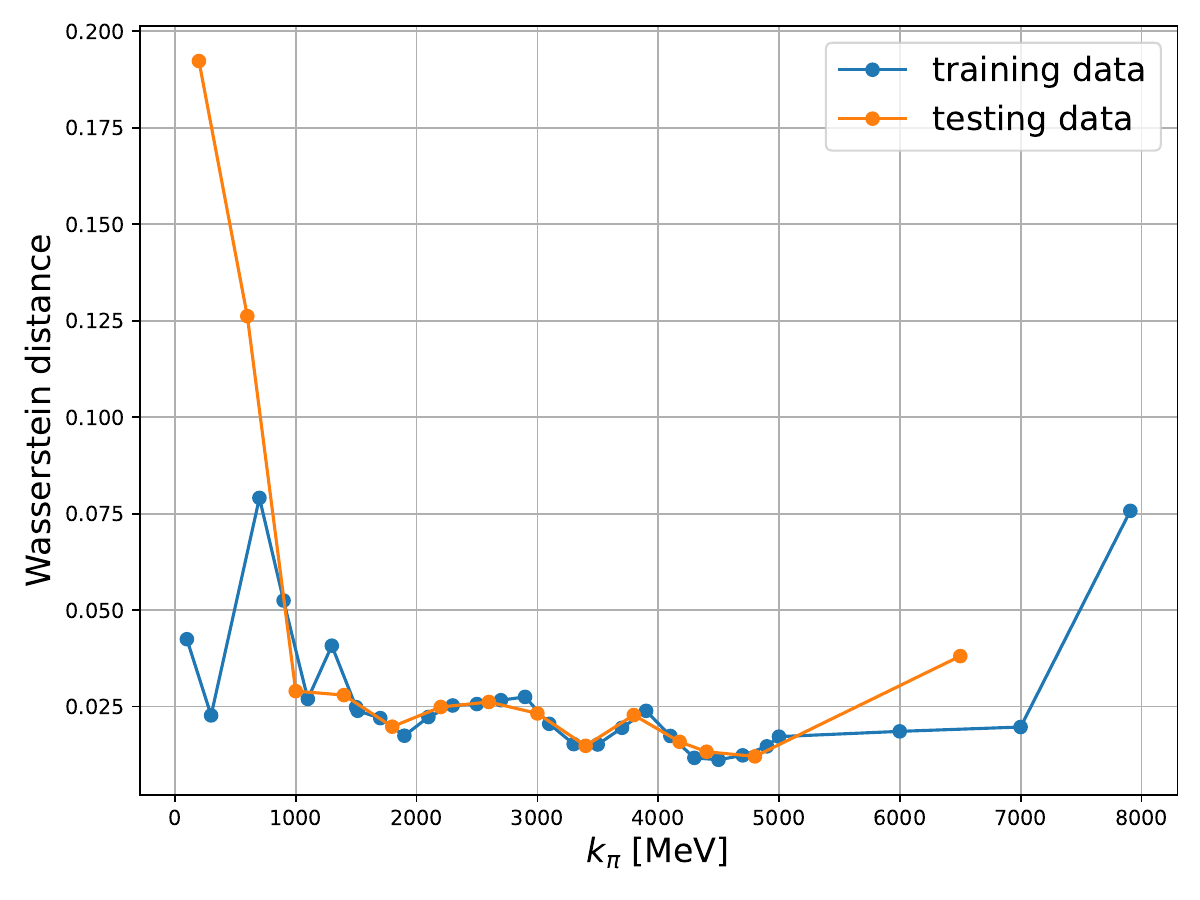}
    \caption{Wassertstein distance between the NF-generated events and the \textsc{GEANT4}-simulated events for different pion energies for the training and testing dataset.}
    \label{fig:res_wd_low}
\end{figure}

We evaluate the performance of the trained normalizing flows using the "Wasserstein distance" (WD)~\cite{Komiske:2019fks,Cai:2021hnn}. This metric is calculated for each variable, comparing the NF-generated events with events simulated using \textsc{GEANT4}. Smaller WD values indicate better performance. Figure~\ref{fig:res_wd_low} presents the WD distances for various pion energies in both the training and testing datasets. The WD spectrum demonstrates that the NF performs exceptionally well within the intermediate range of pion incident energies. However, the NF model struggles to extrapolate the distributions to lower and higher bounds, which could be attributed to fewer sampled incident energies in these regions.

\begin{figure}
    \centering
    \includegraphics[width=0.90\textwidth]{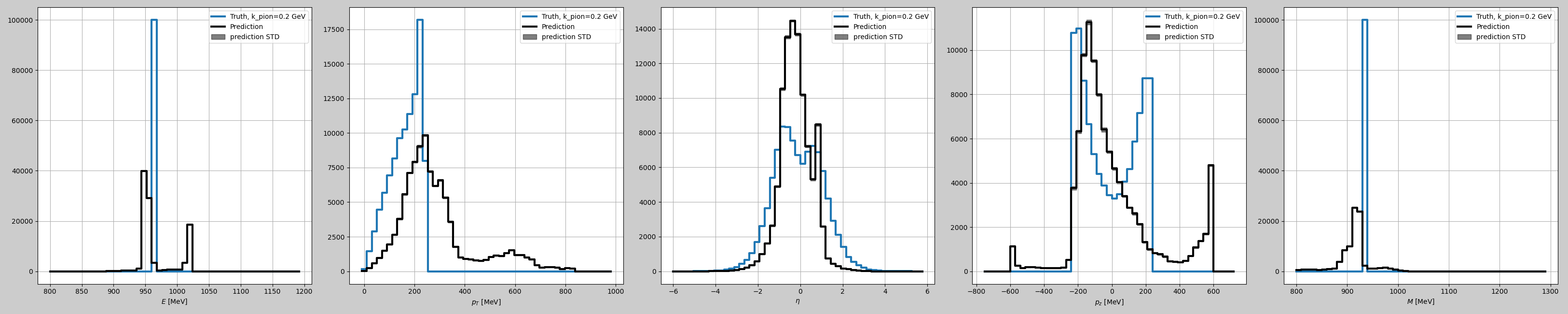}    
    \includegraphics[width=0.90\textwidth]{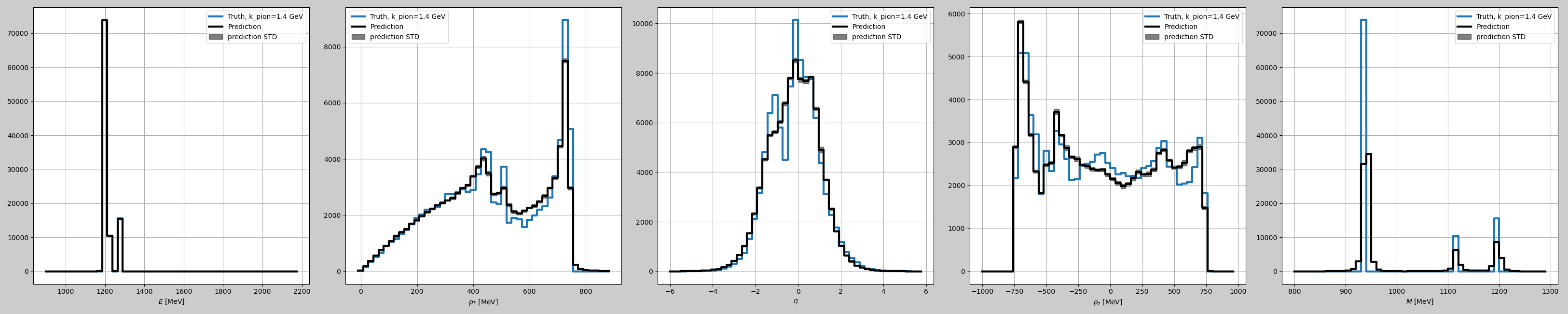}
    \includegraphics[width=0.90\textwidth]{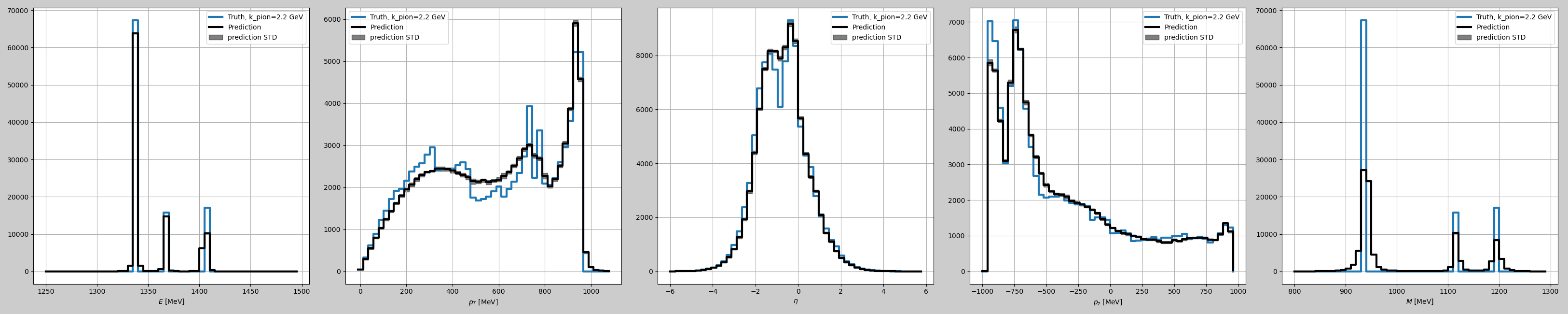}
    \includegraphics[width=0.90\textwidth]{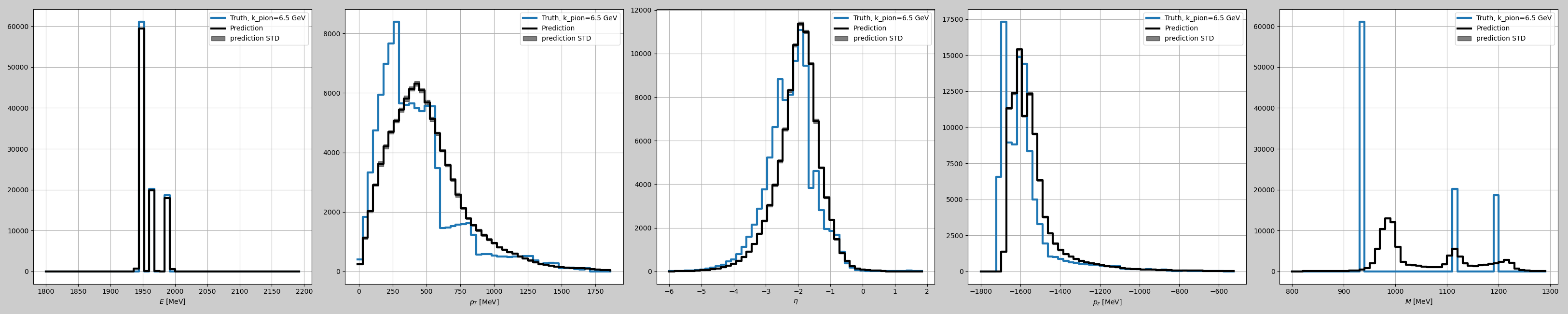}

    \caption{Comparison of energy $E$, transverse momentum \pt, and pseudorapidity $\eta$, $p_z$, and four-vector mass of the leading outgoing particle between \textsc{GEANT4}-simulated events (``Truth'', blue lines) and Normalizing Flow-generated events (``Prediction'', black lines) for incidence pion kinetic energy of 200 MeV, 1.4 GeV, 2.2 GeV, and 6.5 GeV listed from top to bottom. We use the NF model to generate the same amount of events ten times with different seeds. The mean and standard deviations (``Prediction STD'') of the ten distributions are shown.}
    \label{fig:res_lower_energy}
\end{figure}

Figure~\ref{fig:res_lower_energy} compares the distributions generated by the normalizing flow (NF) with \textsc{GEANT4}-simulated ones for incident pion energies that were not part of the training data. It's important to note that the model is specifically trained to predict the four-vector components ($p_x, p_y, p_z$, and $E$) of the particle. To accurately predict the transverse momentum (\pt), the model must capture the correlations between $p_x$ and $p_y$. For incident energies of $k_\pi = 1.4$~GeV and 1.2~GeV, we observe a close agreement between the NF-generated and the \textsc{Geant4}-generated events. However, it becomes evident that more incident energies are necessary for further enhancing the model's performance, particularly in the lower and higher energy regions. 

\begin{figure}
    \centering
    \includegraphics[width=0.8\textwidth]{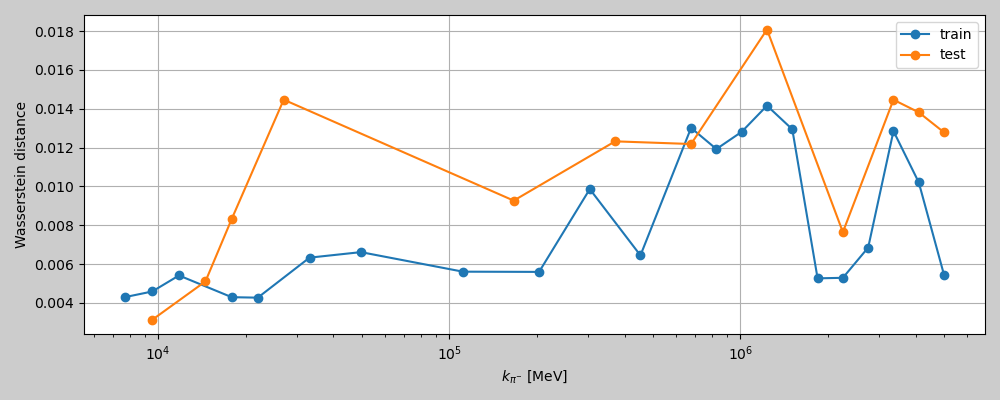}
    \caption{Wassertstein distance between the NF-generated events and the \textsc{GEANT4}-simulated events for different pion energies for the training and testing dataset.}
    \label{fig:res_wd_high}
\end{figure}

In our initial exploration of the data generated at $k_{\pi} > 9$ GeV, we've noticed that the kinematic distributions exhibit a more gradual and continuous variation when only one model, namely the \textsc{FRITIOF} model, is used. Consequently, the normalizing flow can conditionally simulate the data with significantly improved agreement with the true distribution in this region. This improvement is evidenced by the Wasserstein distance observed in Fig.~\ref{fig:res_wd_high}.

\section{Conclusions} \label{sec:conclusions}

Non-perturbative hadronic interactions pose significant physics challenges and formidable computational challenges. Without first-principle theoretical guidance, the primary approach for simulating such interactions is to derive knowledge from experimental measurements, as seen in the implementation of numerous parametric models within the \textsc{Geant4} framework. 

This study represents a promising avenue where Conditional Normalizing Flow (NF) is employed to learn and simulate non-perturbative hadronic interactions based on simulated events. The conditional NF can predict outgoing particle properties with reasonable accuracy in specific energy regions for incident pions.

Nonetheless, there's still more work ahead to achieve a comprehensive simulation of hadronic interactions. The primary challenge lies in generating a variable number of outgoing particles and discrete particle types. The multiplicity of outgoing particles is not solely determined by the incident energy but also by the underlying physics processes. It could be intriguing to develop a physics-informed generative model to address this complexity.

Another significant challenge is constructing a single model covering the entire energy spectrum. In \textsc{Geant4} simulations, particularly in the \textsc{FTFP\_BERT\_ATL} physics list, different models are applied to describe hadronic interactions for incident energies below 9 GeV and those above 10 GeV, with an artifact of mixed models for energies in between. However, it's worth exploring whether a single generative model can simulate hadronic interactions consistently across the entire energy spectrum. Answering this question will require learning from real experimental data rather than solely relying on data generated by \textsc{Geant4}.

\bibliography{graph,main}

\clearpage

\appendix

\end{document}